\documentclass[aps,twocolumn,prl,showpacs,amsmath]{revtex4}
\usepackage[dvips]{graphicx}

\begin{document}

\title{Virial Theorem for Confined Universal Fermi Gases}
\author{J. E. Thomas}
\affiliation{Physics Department, Duke University, Durham, North
Carolina 27708-0305} 

\date{\today}

\begin{abstract}
Optically-trapped two-component Fermi gases near a broad Feshbach
resonance exhibit universal thermodynamics, where the properties
of the gas are independent of the details of the two-body
scattering interactions. We present a global proof that such a
universal gas obeys the virial theorem for {\it any} trapping
potential $U$ and any spin mixture, without assuming either the
local density approximation or harmonic confinement. The total
energy of the gas is given in scale invariant form by
$E=\langle\epsilon\partial U/\partial\epsilon\rangle$, where
$\epsilon$ is an {\it arbitrary} energy scale in terms of which
all length and energy scales that appear in the confining
potential are written. This result enables model-independent
energy measurement in traps that are anharmonic as well as
anisotropic by  observing only the cloud profile, and provides a
consistency check for many-body calculations in the universal
regime.
\end{abstract}

\maketitle

Strongly-interacting Fermi systems, such as neutron matter and
resonantly interacting atomic Fermi gases, can exhibit universal
behavior. These systems satisfy the unitary condition, where the
zero energy scattering length $a$ greatly exceeds the
interparticle spacing, while the range of the scattering potential
is negligible~\cite{BertschChallenge,Baker,Heiselberg}. Unitary
conditions are produced in an optically trapped Fermi
gas~\cite{AmScientist}, by using a magnetic field to tune near a
broad Feshbach resonance~\cite{Stwalley,Tiesinga,Houbiers12},
where strong interactions are observed~\cite{OHaraScience}.
According to the universal hypothesis, unitary systems must
exhibit universal
thermodynamics~\cite{BertschChallenge,Baker,Heiselberg,OHaraScience,HoHighTemp,HoUniversalThermo,BruunUniversal},
since the interparticle spacing sets the only natural length scale
at zero temperature, so that the local properties can be written
generally as functions of the density and temperature. As the
properties of the system are independent of the details of the
interactions, atomic gases can be used to test predictions for
unitary systems in fields well outside atomic physics. Recent
model-independent studies of the entropy versus energy of a
unitary Fermi gas~\cite{LeEntropy} have been used to demonstrate
universal behavior~\cite{DrummondUniversal}. Model-independent
energy measurement is based on the virial theorem, which holds for
a trapped unitary Fermi gas~\cite{ThomasUniversal}.

We have demonstrated both theoretically and experimentally that
the virial theorem  holds for a trapped Fermi gas in the universal
regime at a broad Feshbach resonance~\cite{ThomasUniversal}. The
proof is based on the local density approximation (LDA) with a
local pressure $P=2{\cal E}/3$, which holds for a universal
gas~\cite{HoUniversalThermo}. Here, ${\cal E}$ is the local energy
density (kinetic and interaction energies). Using the balance
between the trapping and pressure forces, one easily obtains the
total energy per particle $E$, which comprises the kinetic energy,
interaction energy, and trap potential energy,
\begin{equation}
E=\left\langle U+\frac{1}{2}\mathbf{x}\cdot\nabla U\right\rangle,
 \label{eq:energymeas}
\end{equation}
where $U$ is the trapping potential. A scalar pressure requires
that $\langle x\partial U/\partial x\rangle=\langle y\partial
U/\partial y\rangle=\langle z\partial U/\partial z\rangle$. In an
anisotropic harmonic trap, this leads to equal energies in all
directions and $E=3m\omega_z^2\langle
z^2\rangle$~\cite{ThomasUniversal}, where $m$ is the atom mass and
$\omega_z$ is the harmonic oscillator frequency for the $z$
direction, which is precisely measurable. In practice, the long
axial direction of the cloud is most easily
imaged~\cite{LeEntropy}. Within the LDA, Eq.~\ref{eq:energymeas}
is readily modified to include corrections arising from
anharmonicity in the trapping potential~\cite{LeEntropy}.
Amazingly, at resonance, a complex many-body Fermi gas, generally
containing condensed (superfluid) atom pairs, noncondensed pairs,
and unpaired atoms, all strongly interacting in the
nonperturbative regime, obeys the same virial theorem as an ideal,
noninteracting gas.

Unfortunately, the local density approximation can break down in
several cases of importance, such as quantum confined systems. In
standing wave optical traps, for example, the harmonic oscillator
frequency for vibrations along the standing wave direction $z$ can
be so large that the axial energy scale $\hbar\omega_z$  is large
compared to the energy in the transverse $x-y$ directions. In
spin-imbalanced mixtures,  phase separation occurs and a surface
tension exists~\cite{MuellerPopImbal,MuellerSurfaceTension}. At
low temperatures, the cloud separates into a superfluid core
containing equal densities of both spins and a region of excess
noninteracting spins outside the core. The shape of the central
core is affected by the unpaired spins, and the pressure is not
necessarily a function of the temperature and local density, since
the spin fractions  vary rapidly in space. The axial dimension is
therefore not necessarily equivalent to the transverse dimensions
of the trap. To avoid the local density approximation, a global
proof of the virial theorem is required.

A global proof of the virial theorem for a universal gas has been
provided by Werner and Castin~\cite{CastinVirial}, who note that
the proof also holds for  anisotropic traps, a result due to
Chevy~\cite{CastinVirial}. Recently, an extremely simple global
proof of the virial theorem for universal gases has been given by
D. T. Son, based on the finite temperature Hellman-Feyman
theorem~\cite{SonVirial}. This proof assumes harmonic confinement,
and is obviously valid for anisotropic harmonic traps and for
spin-imbalanced mixtures. However, it does not allow for
anharmonic trapping potentials. Tan has derived both the pressure
and the virial theorem for a strongly interacting Fermi gas with a
finite scattering length in a generalized confining
potential~\cite{TanVirial}. The results show that the correction
is $\propto 1/a$, as confirmed by Braaten and
Platter~\cite{BraatenVirial}, who obtain the same result as Tan by
using the Hellman-Feynman theorem. One expects that the virial
theorem also will hold for a universal gas with three body
resonances, as discussed recently by Nishida, Son, and
Tan~\cite{Nishida3Body}.

We now provide a simple global proof of the virial theorem in the
universal regime valid for any spin mixture and {\it any} trapping
potential. The result enables a general method for precisely
measuring the total energy of a universal gas in anharmonic
trapping potentials. Our proof follows closely the method employed
by Son~\cite{SonVirial}, and yields a general, global,
scale-invariant relation between the energy and the trapping
potential, without assuming harmonic confinement. This permits the
determination of small corrections for anharmonicity in the
trapping potential, which is essential for precision measurements
of the energy.

First, consider a Hamiltonian that is a function of a parameter
$\lambda$, i.e., $H(\lambda)$. Then, the expectation value of
$H(\lambda)$ with the $n^{th}$ eigenstate yields $\langle
n,\lambda|H(\lambda)|n,\lambda\rangle = E_n(\lambda)$, where
$E_n(\lambda)$ is the eigenvalue. Taking the derivative of this
equation with respect to $\lambda$ and noting that the states
remain normalized as $\lambda$ varies yields the Hellman-Feynman
theorem for one eigenstate, $\partial
E_n(\lambda)/\partial\lambda=\langle n,\lambda|\partial
H(\lambda)/\partial\lambda|n,\lambda\rangle$.

We assume that the evaporation rate from a trap of finite depth is
sufficiently slow that the system is in quasi equilibrium, so that
 the probability for each many-body quantum state is determined
 by a Boltzmann factor for a fixed total number of atoms $N$.
 Then, following Son~\cite{SonVirial}, we consider the free energy,
$F=-k_BT\ln Z$, where $Z$ is the partition function,
$Z=\Sigma_n\exp[-\beta\,E_n(\lambda)]$, and $\beta = 1/(k_BT)$.
Differentiating $F$ with respect to $\lambda$ and using the
Hellman-Feynman result for each eigenstate easily yields the
finite temperature Hellman-Feynman theorem for any parameter
$\lambda$,
\begin{equation}
\frac{\partial F(\lambda)}{\partial\lambda}=\left\langle
\frac{\partial H(\lambda)}{\partial\lambda}\right\rangle.
\label{eq:FiniteTHFthm}
\end{equation}

The Hamiltonian $H$ generally comprises the total kinetic energy
operator, which introduces the parameter $\hbar^2/m$ with $m$ the
atom mass, the interaction potential, which introduces the s-wave
scattering length, and the trapping potential energy $U=\sum_i
U_{trap}(\mathbf{r}_i)$ with $\mathbf{r}_i$ the position of the
$i^{th}$ atom. In the universal regime, the details of the
resonant two-body interaction potential cannot appear in the
properties of the gas: Compared to the interparticle spacing, the
zero-energy s-wave scattering length  $a$ for opposite spin atoms
is infinite, and the range $R$ of the potential is zero. The
corresponding energy scales are $\hbar^2/(ma^2)$ and
$\hbar^2/(mR^2)$, which are zero and infinity, respectively. For
identical spin atoms, the scattering length is zero. Hence, for
both resonantly interacting and noninteracting atoms, {\em only}
the trapping potential $U$ introduces natural energy scales.

We choose {\em one} arbitrary energy scale $\equiv\epsilon$, for
example, 1 joule, and write all other energy scales  $\epsilon_i$
that appear in the trapping potential in terms of $\epsilon$,
i.e., $\epsilon_i=\alpha_i\epsilon$, as described further below.
Now, let $\lambda =\epsilon$ in Eq.~\ref{eq:FiniteTHFthm}, and
note that $\langle
\partial H/\partial\epsilon\rangle =\langle\partial
U/\partial\epsilon\rangle$, since only the trap potential can
contain the energy scale $\epsilon$. Then, $\partial
F(\epsilon)/\partial\epsilon=\langle\partial
U/\partial\epsilon\rangle$.

In the universal regime, the free energy $F$ can be written in
terms of the chosen energy scale, and it must be of the form
$F=\epsilon\,f(k_BT/\epsilon)$, where $f$ is a dimensionless
function of the temperature $T$ . Note that we suppress all other
variables, such as the total atom number $N$, the spin fractions,
$N_1/N$, the $\alpha_i$ etc, which are held constant.
Differentiating $F$ with respect to $T$, we obtain the entropy
$S=-\partial F/\partial T=-k_B\,f'(k_BT/\epsilon)$, where $f'$
denotes the derivative of $f$ with respect to its argument. Then,
$\epsilon\partial F/\partial\epsilon=F-k_BT\,f'=F+TS=NE$, where
$E$ is the total energy per particle. Hence, we obtain a simple,
general, and global  relation between the total energy per
particle and the trapping potential,
\begin{equation}
E=\left\langle\epsilon\frac{\partial
U}{\partial\epsilon}\right\rangle, \label{eq:universalenergy}
\end{equation}
where the brackets denote a statistical and quantum average with
the density operator. Eq.~\ref{eq:universalenergy} is obviously
independent of the energy scale $\epsilon$.

As a simple example, consider a power law trapping potential of
the form $U_n=U_0(r/a_{trap})^n$, where $r$ is the position of an
atom of mass $m$. To obtain normalizable states, we assume a
confining potential, where $U_0>0$ for $n>0$ and $U_0<0$ for
$n<0$. Let $U_0=\alpha_1\epsilon$ and
$\hbar^2/(ma_{trap}^2)=\alpha_2\epsilon$. Then,
$U_n=\epsilon^{1+n/2}\alpha_1\alpha_2^{n/2}(mr^2/\hbar^2)^{n/2}$.
Eq.~\ref{eq:universalenergy} immediately yields,
\begin{equation}
E=(n/2+1)\langle U_n\rangle, \label{eq:virialpowerlaw}
\end{equation}
which is the standard result.

Now consider an isotropic gaussian trap, for which the single
particle trap potential energy is $U_{trap}=U_0[1-\exp(-
r^2/a^2_{trap})]$. Writing the energy scales associated with the
trap depth and 1/e width as above, we have,
$$U_{trap}=\alpha_1\epsilon\,\left[1-\exp\left(-\alpha_2\epsilon\,\frac{mr^2}{\hbar^2}\right)\right].$$

For small displacements, the trap potential is harmonic:
$U_{trap}=\alpha_1\alpha_2\epsilon^2\,\frac{mr^2}{\hbar^2}=m\bar{\omega}^2r^2/2$
with $m\bar{\omega}^2=2U_0/a^2_{trap}$.
Eq.~\ref{eq:universalenergy} immediately yields the virial theorem
result, $E/N=\langle m\bar{\omega}^2r^2\rangle$. Note that the
oscillator energy is
$\hbar\bar{\omega}=\sqrt{2\lambda_1\lambda_2}\epsilon$ and
therefore introduces no new energy scales.

Applying Eq.~\ref{eq:universalenergy} to the gaussian potential,
and keeping up to lowest order in the anharmonic corrections, we
obtain the energy per particle
$$E=\left\langle
m\bar{\omega}^2r^2-\frac{3}{8}\frac{(m\bar{\omega}^2r^2)^2}{U_0}\right\rangle.$$
Here the averages are determined experimentally by measuring the
cloud profiles for {\it both} spin states in spin-imbalanced
mixtures. This reproduces our previous expression for the
anharmonic correction~\cite{LeEntropy}, which was obtained
assuming the local density approximation with a scalar pressure.
For an anisotropic gaussian trapping potential, we simply replace
$\bar{\omega}^2r^2$ by $\omega_x^2x^2+\omega_y^2y^2+\omega_z^2z^2$
and use $\bar{\omega}\equiv(\omega_x\omega_y\omega_z)^{1/3}$. The
proof goes through unchanged.

For a nearly harmonic trap with cylindrical symmetry in $x$ and
$y$, the energy is readily determined from the two dimensional
$x-z$ cloud profile. The energy per particle is
\begin{equation}
E=2m\omega_x^2\langle x^2\rangle+m\omega_z^2\langle z^2\rangle.
\label{eq:globalEHO}
\end{equation}
Here we do {\it not} assume that the axial and transverse
contributions are identical, as required in the local density
approximation, since the local density approximation may not hold.

In general, the quartic terms for the  trapping potential can be
measured. This is accomplished by determining the parametric
resonance frequencies  and the trap average values of $\langle
z^4\rangle$, $\langle x^4\rangle$, and $\langle z^2x^2\rangle$
from the cloud profiles for an ideal Fermi gas tuned near the zero
crossing  where the scattering length is zero, i.e., near 528 G
for $^6$Li. For a trap with near cylindrical symmetry, we expect
that the total potential has symmetry under $\omega_x
x\leftrightarrow \omega_y y$. For a given atom number, the energy
is varied by release and recapture as described
previously~\cite{LeEntropy}. The dependence of the frequency on
the quartic averages of the spatial coordinates, which are
determined from the cloud profiles, will provide the appropriate
anharmonic corrections to Eq.~\ref{eq:globalEHO}.

This research was supported by the Physics Divisions of the Army
Research Office  and  the National Science Foundation,  and the
Chemical Sciences, Geosciences and Biosciences Division of the
Office of Basic Energy Sciences, Office of Science, U. S.
Department of Energy.

%\bibliography{FermiGas308}

\end{document}